\documentclass{iau} 
\usepackage{graphicx,natbib}

\let\boldgrk=\gkvecten
\let\boldgrksc=\gkvecseven

\def\gkthing#1{{\mathchoice%
	{\hbox{{\boldgrk\char#1}}}
	{\hbox{{\boldgrk\char#1}}}
	{\hbox{{\boldgrksc\char#1}}}
	{\hbox{{\boldgrksc\char#1}}}}}

\def\vtheta{\gkthing{18}}

{\newif\ifnotend
\notendtrue
\def\veclist{ABCDEFGHIJKLMNOPQRSTUVWXYZabcdefghijklmnopqrstuvwxyz.}
\def\top#1#2.{#1}
\def\tail#1#2.{#2.}
\loop\expandafter\xdef\csname v\expandafter\top\veclist\endcsname%
{{\noexpand\bf\expandafter\top\veclist}}
\edef\veclist{\expandafter\tail\veclist}
\if\veclist.\notendfalse\fi\ifnotend\repeat}
\def\spose#1{\hbox to 0pt{#1\hss}}
\def\lta{\mathrel{\spose{\lower 3pt\hbox{$\mathchar"218$}}
     \raise 2.0pt\hbox{$\mathchar"13C$}}}
\def\d{{\rm d}}\def\e{{\rm e}}\def\i{{\rm i}}
\def\Rc{R_{\rm c}}\def\Rd{R_{\rm d}}\def\Vc{v_{\rm c}}
\def\Myr{\,{\rm Myr}}\def\kpc{\,{\rm kpc}}\def\kms{\,{\rm km\,s}^{-1}}
\def\msun{\, M_\odot}\def\pc{\,{\rm pc}}
\def\fracj#1#2{{\textstyle{#1\over#2}}}

\title[Self-consistent modelling of our Galaxy] 
{Self-consistent modelling of our Galaxy with Gaia data}

\author[James Binney]   
{James Binney}

\affiliation{Rudolf Peierls Centre for Theoretical Physics, University of
Oxford, 1 Keble Road, Oxford OX1 3NP, UK \\ email: {\tt
binney@thphys.ox.ac.uk}}

\pubyear{2017}
\volume{330}  
\jname{Astrometry and Astrophysics in the Gaia sky}
\editors{Alejandra Recio-Blanco, Patrick de Laverny, Anthony Brown and, Timo
Prusti eds.}
\begin{document}

\maketitle

\begin{abstract}
Galaxy models are fundamental to exploiting surveys of our Galaxy. There is
now a significant body of work on axisymmetric models. A model can be
defined by giving the DF of each major class of stars and of dark matter. Then
the self-consistent gravitational potential is determined. Other modelling
techniques are briefly considered before an overview of 
some early work on
non-axisymmetric models.
\end{abstract}

\firstsection 
\section{The Standard Galaxy}

In 2027 our understanding of the Galaxy will be encapsulated in a software
tool that I'll call the Standard Galaxy (SG). The SG, like a Wiki page, will
always be a work in progress.  When a survey is planned, the SG will be used
to simulate the survey's contents in light of its selection function (SF). When the survey is completed, the SG
will be updated by maximising the likelihood of the new data with respect to
the SG's parameters and the priors from earlier surveys. The SG will describe
what's actually out there, which each survey sees only partially.

\subsection{What's in the SG?}

\begin{itemize}

\item The SG will model the distribution in $(\vx,\vv)$ of
many types of stars:

\begin{itemize}
\item O, B, A, F, G K, M dwarfs, Cepheid variables, RR Lyrae stars, BHB
stars, red clump stars, white dwarfs, neutron stars, \dots

\item Most stellar types will be subdivided by age, [Fe/H] and [$\alpha$/Fe]
and some will be subdivided by other abundance ratios.

\end{itemize}

\item The SG will also specify the distribution in $(\vx,\vv)$ of dark matter (DM).

\item The SG will include 3-dimensional models of HI, H$_2$, the density of H$^+$
and possibly other chemical species. It will include a model of the  interstellar  velocity
field, $\vv(\vx)$.

\end{itemize}

\subsection{The SG and Jeans' theorem}

DM plays an essential role in Galactic structure. But we cannot (yet)
actually see it, and there is no guarantee that DM particles will have been
detected by 2027. Indeed, even if current efforts to detect DM particles
underground bear fruit, measurements of their density and velocity
distribution on Earth will add only moderately to what we have already
discovered about DM by modelling its contribution to the Galaxy's
gravitational field.

The process by which we currently constrain DM by modelling its gravitational
field is completely reliant on the assumption that the Galaxy is close to
statistical equilibrium. If we drop the assumption that the Galaxy looked
pretty much the same $200\Myr$ ago, and will look essentially the same
$200\Myr$ hence, we can infer nothing about the Galaxy's gravitational field
from the kinematics of stars and gas, because without this assumption of
statistical equilibrium any current distribution of matter in phase space is
consistent with any gravitational field.

From our reliance on statistical equilibrium to track $\sim95\%$ of the
Galaxy's mass, it follows that the foundations of the SG will rest on an
equilibrium model of the Galaxy. In reality, the Galaxy is not in
equilibrium, even in the rotating frame of the bar, because the stellar halo
hosts tidal streams and the disc displays ephemeral spiral arms and a warp.
These non-stationary phenomena play key roles in the Galaxy's evolution
\citep[e.g][]{Aumer2017} and promise to be valuable probes of the Galaxy's
assembly history \citep[e.g.][]{Erkal}, but they must be excluded from the
basic model, which will inevitably be an equilibrium model. Only
after its construction will it be decorated with spiral arms, warps and streams.

The natural way to construct an equilibrium model is to exploit Jeans'
theorem: the distribution function (DF) of an equilibrium stellar system may
be presumed to depend on $(\vx,\vv)$ through the constants of stellar motion
$I_1(\vx,\vv),I_2(\vx,\vv),\ldots$.
Since any function of constants of motion is itself a
constant of motion, we have infinite freedom in the choice of the constants
of motion that we use as arguments of the DF $f(\vI)$. There is, however,
only one rational choice: the action integrals $J_i$
\citep[e.g.][]{BinneyMcM}.
Until recently the use of angle-action variables in galactic dynamics was
very limited because we lacked effective ways to convert action-angle
coordinates to and from conventional phase-space coordinates. Several
numerical schemes for making these transformations are now available
\citep{SandersActions}. The ``St\"ackel Fudge'' \citep{BinneySF} provides the
most widely used mapping from $(\vx,\vv)$ to $(\vtheta,\vJ)$ while Torus Mapping
\citep{BinneyMcM} provides the best inverse map. In particular cases the
existing tools do have limitations, so further work is required in this area.

\subsection{How to assemble a model}

First for each component one wishes to consider, one chooses either a fixed
density distribution $\rho(\vx)$ or a DF $f(\vJ)$. The normalisation of
either $\rho$ of $f$ is adjusted to ensure that the component's mass is
reasonable. Then one makes a guess at the gravitational potential $\Phi(\vx)$
of the final model, and using this potential, the St\"ackel Fudge and the DFs
one evaluates the model's density by integrating over velocities at each
point of a suitable spatial grid. Then one computes the resulting potential
$\Phi(\vx)$ and uses this potential instead of the originally guessed
potential to re-compute the density at the grid points. This cycle of
computing the density using one potential and from it deriving a new
potential converges after 4 to 5 iterations \citep{BinneyHenon}. From this
potential and the DFs one can predict {\it any} observable. The parameters in
each component's DF are adjusted to optimise the fit between the model and
observational data \citep{BinneyPiffl}.

Given the long tradition of including energy $E=\fracj12v^2+\Phi$ in the
arguments of the DF, it's worth noting that model construction proceeds
smoothly as just described only if $E$ is excluded from the arguments of $f$.
When $E$ is included it's hard to converge on the correct potential, and
harder still to ensure that each component has an observationally motivated
mass.

\subsection{Choice of DFs}

It turns out that simple analytic functions $f(\vJ)$ generate models that
closely resemble familiar models that are defined by simple functional forms
of the density $\rho(\vx)$. The isochrone sphere can be exactly generated
from an analytic $f(\vJ)$ \citep{BinneyHenon}. \cite{Posti} showed that the
Hernquist, Jafffe and NFW \citep{NFW} spheres can be generated to high
precision by simple analytic functions $f(\vJ)$. \cite{xSanders} have given
an analytic $f(\vJ)$ which generates a good approximation to a Plummer model.
By tweaking the given forms of $f(\vJ)$, a spherical model can be endowed with
either tangential or radial anisotropy, and it can be flattened or made prolate
by the anisotropy \citep{BinneyHenon,BinneyPiffl}.

It is useful to break the DF into a parts $f_+(\vJ)$ and $f_-(\vJ)$ that are
even and odd in $J_\phi$, respectively. The part $f_-$ does not
contribute to the density but instead controls the model's rotation.
Consequently, a model's rotation can be readily adjusted to match
observational data after its density distribution has been perfected.

\cite{BinneyMcMillan} introduced the ``quasi-isothermal'' DF for discs:
\[
f_+(\vJ)={\Sigma_0\Omega\over\kappa^2}\exp(-\Rc/\Rd){\exp(-\kappa
J_r/\sigma_r^2)\over\sigma_r^2\kappa^{-1}}{\exp(-\nu
J_z/\sigma_z^2)\over\sigma_z^2\nu^{-1}}
\]
 as the simplest DF that creates a plausible disc. Here $\Sigma_0$ and $\Rd$
are constants that, respectively, set the disc's mass and approximate scale
length, while $\Rc,\Omega,\kappa,\nu,\sigma_r$ and $\sigma_z$ are all
functions of $J_\phi$. Specifically, $\Rc$, $\Omega$, $\kappa$ and $\nu$
should be the radius, angular velocity, in-plane and vertical epicycle
frequencies of a circular orbit of angular momentum $J_\phi$ in some
potential that is similar to that of the Galaxy. The velocity-dispersion
parameters $\sigma$ should be decreasing functions of $J_\phi$ and the normal
hypothesis is
\[
\sigma_i=\sigma_{i0}\exp\left[-{\Rc-R_0\over R_{\sigma i}}\right]
\]
 where $\sigma_{i0}$ is a constant that sets the $i$th velocity dispersion at the
Sun and $R_{\sigma i}$ is a constant that determines how rapidly $\langle
v_i^2\rangle$ declines with distance from the Galactic centre. In a realistic
Galactic potential  the
quasi-isothermal DF produces a disc with an approximately exponential surface
density $\Sigma(R)\simeq\Sigma_0\e^{-R/\Rd}$ and a vertical density profile
that is sub-exponential in the sense that $|\d\log\rho/\d z|$ is a slowly
increasing function of distance $z$ from
the plane. However,  a vertical density profile in which
$|\d\log\rho/\d z|$ decreases with $z$ as is observed
\citep{GilmoreReid,Juric08} emerges naturally when one models the observed secular
increase in $\langle v_z^2\rangle$ with age $\tau$ by modelling each coeval
population  of stars by a quasi-isothermal with, for example
\[
\sigma_{i0}(\tau)=\sigma_{i*}\left({\tau+\tau_1\over\tau_{\rm m}+\tau_1}\right)^\beta,
\]
 where $\sigma_{i*}$ and $\tau_{\rm m}$ are the  velocity dispersion and age
of the oldest disc stars, $\sigma_1$ determines the velocity dispersion of
these stars at birth and $\beta\sim0.5$ controls how the velocity dispersion
increases with age.

To date $f_-$ has been taken to be 
\[
f_-(\vJ)=\tanh(J_\phi/J_0)f_+(\vJ),
\]
 where $J_0$ is a constant. This ansatz eliminates counter-rotating stars at
angular momenta significantly larger than $J_0$, which is assumed to be much
smaller that $R_0\Vc(R_0)$.

\section{What's been done so far}

Several papers have fitted models based on DFs $f(\vJ)$ to data by evaluating
moments of the DF in an assumed gravitational potential $\Phi(\vx)$. Adopting
a plausible $\Phi(\vx)$ rather than solving for the self-consistent potential
saves a great many CPU cycles because it is then necessary to compute moments
only at locations for which we have data, rather than throughout the vast
extent of the Galaxy (which extends out to $>100\kpc$). Moreover, with
$\Phi(\vx)$ assumed, we only need the  moments of components for which we
have data. In particular, we don't need to compute moments for the Galaxy's
principal component, dark matter.

\cite{Binney2010} fitted a disc DF to data from the Geneva-Copenhagen survey
(GCS) \citep{Nordstrom,Holmberg}. The most significant finding of this paper
was that the Sun's peculiar velocity $V_\odot$ needed to be revised upwards
from $5.2\kms$ to $\sim11\kms$. \cite{SchoenrichBD} explained how the
standard extraction of $V_\odot$ from the Hipparcos data had been undermined
by the metallicity gradient in the disc. \cite{Binney2012} upgraded his
earlier work by using an improved the quasi-isothermal DF (hereafter the
``2012 DF'') and the just
introduced St\"ackel Fudge rather than the adiabatic approximation to compute
actions. \cite{BinneyRAVE} showed that the 2012 DF had great
success in predicting the data from the RAVE survey, which reaches distances
in excess of $2\kpc$ whereas the GCS is essentially confined to
distances $<0.1\kpc$. The extent to which the 2012 DF captures the strong
non-Gaussianity of the velocity distributions in $v_\phi$ and $v_z$ is
remarkable.

\cite{SandersEDF} proposed an extended DF (EDF) in which [Fe/H] appears
alongside $\vJ$ as an argument. Since stellar age already appeared in the
2012 DF as a nuisance parameter, with [Fe/H] added to the argument list it
became possible to employ stellar isochrones to compute the probability that
a star in the model would be included in a given survey. Hence
\cite{SandersEDF} were able to take properly into account survey SFs, which
\cite{Binney2012} and \cite{BinneyRAVE} had neglected to do.  They showed
that when the SF of the GCS is taken into account, significantly larger
values of $\sigma_{i*}$ are required because the GCS is strongly biased
towards young stars, so the stars picked up in the survey have atypically
small random velocities.

By fitting an EDF to a sample of SDSS K giants \cite{Dasa} found evidence for
two subpopulations. The EDF \cite{Dasb} fitted to BHB stars showed the older
stars to be more tightly confined in action space than the younger stars.
\cite{BinneyWong} explored DFs for the Galaxy's disc and halo globular
clusters. This exercise revealed that a featureless DF $f(\vJ)$ is liable to
generate a density distribution $\rho(\vx)$ on which  the gravitational
potential has
imprinted a feature associated with its transition around $r\sim10\kpc$ from
disc- to halo-domination.
 
The moments yielded by a given DF depend on the adopted potential $\Phi$.
Consequently, $\Phi$ can be constrained by fitting to data the real-space and
velocity-space distributions obtained by a particular pair $(\Phi,f)$.
\cite{BovyRix} computed the likelihoods of 43 groups of stars over a
5-dimensional grid in $(\Phi,f)$ space.  Each stellar group comprised the
$\sim400$ G dwarfs from the SEGUE survey that lie in a cell in the
([Fe/H],[$\alpha$/Fe]) plane. The DF was constrained to be a quasi-isothermal and
$\Phi$ was  generated by a spherical bulge and dark halo plus
double-exponential stellar and gas discs.  The bulge was a fixed Hernquist
model and the dark halo was a power-law model, so its free parameters were
its logarithmic slope and its local density. The mass and scale lengths of
the  stellar disc were free parameters.
The likelihood of each stellar
group was computed over a 5-dimensional grid in $(\Phi,f)$ space.
Unfortunately, the likelihood distributions in $(\Phi,f)$ space of the
different populations were not mutually consistent. This is not surprising
since the phase-space distribution of stars of a given chemical composition
cannot be well modelled by a quasi-isothermal. In particular, stars with low
$[\alpha$/Fe] and low [Fe/H], being relatively young low-metallicity stars, have
to be confined to an annulus centred beyond $R_0$.

\cite{PifflRAVE} took a different approach to choosing a $(\Phi,f)$ pair, and
they exploited the kinematics of $\sim200\,000$ giant stars from the RAVE
survey rather than $\sim17\,000$ SEGUE stars. They computed a $\chi^2$
statistic for the fit provided by $(\Phi,f)$ to both the RAVE kinematics and
the density of stars $\rho(z)$ extracted from SDSS star counts by
\cite{Juric08} and two diagnostics of the Galaxy's circular-speed curve:
terminal velocities measured from CO and HI data, and astrometry of maser
sources. They homed in on the best $(\Phi,f)$ pair by first adopting a local
dark-halo density $\rho_{\rm DM}$, finding the disc that then best reproduces
the diagnostics of the circular speed and the kinematics of RAVE giants and
then computing the resulting vertical density profile of the disc. If the
initially assumed $\rho_{\rm DM}$ is small, a massive disc is required to
match constraints on the rotation curve, and then given the RAVE kinematics
the disc is too thin to match the SDSS $\rho(z)$. When $\rho_{\rm
DM}\simeq(0.013\pm0.002)\msun\pc^{-3}$ all data are fitted quite nicely. The
principal uncertainty is the shape of the dark halo: a more oblate dark halo
requires a less massive disc to complement it. It turns out that the mass of
the dark halo at $R<R_0$ is almost independent of the halo's axis ratio.

\cite{PifflPB} first took the major step of specifying the dark halo by a DF
$f(\vJ)$ rather than a parametrised density distribution. Once this step is
taken, it no longer makes sense to use a parametrised potential $\Phi$ and
the model is completely specified through the self-consistency condition by
the DFs of its constituent populations.  For the disc, Piffl et al.\ adopted
the DF fitted by \cite{PifflRAVE}, while for the dark halo they chose the
DF that would in isolation self-consistently generate the NFW density profile
determined by \cite{PifflRAVE}. When the self-consistent model implied by
these two DFs was constructed, its circular-speed curve proved inconsistent with
the data at $r\lta3\kpc$ because the dark halo was pulled inwards and towards
the plane by the gravitational field of the disc.

\cite{BinneyPiffl} took the obvious next step: search the space $(f_{\rm
disc},f_{\rm DM})$ for a pair of DFs that self-consistently generate a
model that is consistent with all the data assembled by \cite{PifflRAVE}. In
this search the functional forms of the DFs were the same as those adopted by
\cite{PifflPB} but the parameters were free. A model that satisfied the
observational constraints was found. It did however, have a remarkably large
disc scale-length ($3.7\kpc$) and it did not produce as high an optical depth
to microlensing bulge stars as microlensing surveys require \citep{SumiPenny}. The
reason for these short-comings was clear: in order to keep the circular speed
at $R\sim3\kpc$ within the observational upper limit, stars had been shifted
outwards by increasing $\Rd$, leaving dark matter, which does not cause
microlensing, the dominant mass density at small radii.

This finding establishes a tension between (i) the local density of DM that one
deduces from $\Vc(R_0)$ and the kinematics and thickness of the disc, (ii) the
assumption that DM has been adiabatically compressed by the gravitational
field of slowly inserted baryons, and (iii) the density of stars at $R\lta3\kpc$
required by the microlensing data. The clear next step is to drop the
assumption that DM has responded adiabatically to the insertion of baryons.

The NFW density profile implies that $f_{\rm DM}$ diverges as $\vJ\to0$,
because the divergence of $\rho_{\rm DM}(r)$ as $r\to0$ is accompanied by
decreasing velocity dispersion. This prediction of simulations of
cosmological clustering in the absence of baryons must be a consequence of
the assumed extremely large density of DM at high redshift. The very high
phase-space density of DM at the centre of an NFW halo will be drastically
lowered when a passing clump of baryons imparts even a tiny velocity kick to
DM particles. Once the phase-space density of DM has been reduced in this
way, by Liouville's theorem it cannot be increased. Moreover, the efficient
scattering of particles in a confined region of phase space will drive the
phase-space density to a constant, which is the state of maximum entropy
subject to constrained mean phase-space density. So \cite{ColeB} took the
view that a probable $f_{\rm DM}(\vJ)$ is one that tends to a constant as
$\vJ\to0$, where scattering by baryons has been efficient, and tends to the
NFW $f(\vJ)$ at high energies, where scattering has been unimportant. The
transition between these two asymptotic regimes must be consistent with
conservation of DM by scattering.  \cite{ColeB} devised a form for $f_{\rm
DM}(\vJ)$ that is consistent with these principles and successfully searched
the space $(f_{\rm disc},f_{\rm DM})$ for models consistent with the P14 data
plus the microlensing constraints. Their new $f_{\rm DM}(\vJ)$ has an
additional parameter $h_0$, which specifies the transition between the regime
of approximately constant $f_{\rm DM}$ and the regime in which $f_{\rm
DM}\sim f_{\rm NFW}$. In the absence of baryons their favoured
$h_0\sim150\kpc\kms$ implies a DM core radius $\sim3\kpc$, but the dark
halo doesn't have a core in 
presence of the baryons. 

Through the sequence of models of increasing sophistication just described,
the local density of DM has changed very little from the value $\rho_{\rm
DM}\simeq0.013\msun\pc^{-3}$ determined by \cite{PifflRAVE}.

\subsection{Models not based on actions}

Most models of external galaxies are based either on the Jeans equations or  on the technique introduced by
\cite{Schwarzschild} \citep{vdVen}. The best current models of the
Galactic bar \citep{Portail} use the ``made-to-measure'' variant  of
Schwarzschild's technique \citep{SyerTremaine,DeLorenzi}. Jeans models
certainly lack the rigour and flexibility required to do justice to Galactic
data. In Scharzschild and made-to-measure models the
initial conditions of numerically integrated orbits play the role of
constants of motion and weights assigned to orbits play the role of the value
taken by the DF on an orbit. It is relatively simple to fit such models to
observational data, and neither deviations from axisymmetry nor resonant
trapping are problematic. The major differences with $f(\vJ)$ modelling are
(i) the DFs have vastly more parameters than  a typical $f(\vJ)$, and (ii)
the orbit labels they use are complex and devoid of physical meaning. 

I see several reasons why  these models are unlikely to rise to the challenge posed by
data in the era of Gaia. First, such a model is cumbersome because it is
specified by millions of weights with low individual information content.
Consequently, it's hard to compare models -- two models with identical
physical content will have no two parameters the same because each model will
sample phase space in a different way. Moreover, the lack of a systematic
scheme for labelling orbits that's invariant under changes in the
gravitational potential means that it's hard to refine a model by subdividing
components into sub-groups. For example, as data on the age distributions of
giants accumulate, one will want to subdivide stars on the giant branch into
age cohorts. Such refinements will invariably be associated with a change in
the self-consistent potential and it's important to be able to identify
orbits before and after such a change.

An additional issue with models of the Schwarzschild type is that it is
difficult to include DM in the modelling process. One issue is that the quantity
and spatial extent of DM necessitates a huge increase in the particle count.
A more profound issue is that {\it no} observational data directly constrain
DM -- all constraints come indirectly through observational constraints on
visible matter. It is not clear that current techniques can assign weights to
DM particles.

It will never be possible to fit a standard N-body model to the exquisite
data for our Galaxy, but generic N-body models have a huge role to play
because they uniquely enable us to model from first principles  evolutionary
processes such as secular heating and radial migration
\citep[e.g.][]{AumerBS,Aumer2017}.

\section{What next?}

Not only is our Galaxy  barred, but numerical simulations of the secular
growth of a disc within a dark halo show that spontaneously generated, transient non-axisymmetries
account very naturally for the structure of the thin disc
\citep{AumerBS,Aumer2017}. Moreover, it now seems likely that the bar's
corotation resonance lies as far out as $R\simeq6\kpc$, so the kinematics of
solar-neighbourhood stars are  significantly affected by the bar. Hence it is
essential to regard axisymmetric models as spring-boards from which to
progress to models that include both a bar and spiral structure.

As noted above, angle-action variables provide the natural arena for
perturbation theory, and a major attraction of models built around analytic
DFs $f(\vJ)$ is the ease with which we can perturb them.
The impact of a perturbing potential $\Phi_1(\vtheta,\vJ)$ on a model with
DF $f(\vJ)$ can be computed by linearising the Boltzmann equation:
$f_0(\vJ)\to f_0(\vJ)+f_1(\vtheta,\vJ,t)$. With  $[a,b]$ denoting the
Poisson bracket, we have
\[
{\p f_1\over\p t}+[f_0,\Phi_1]+[f_1,\Phi_0]=0.
\]
 This equation is readily solved once one knows the Fourier expansion
$\Phi_1=\sum_\vn \phi_\vn(\vJ)\e^{\i\vn\cdot\vtheta}$ of the perturbing
potential.  \cite{Monari} have used this approach to compute the response of
thin-disc stars at the bar's outer Lindblad resonance (OLR), which
\cite{Dehnen} argued lies just beyond $R_0$.

\begin{figure}
\begin{center}
\includegraphics[width=.5\hsize]{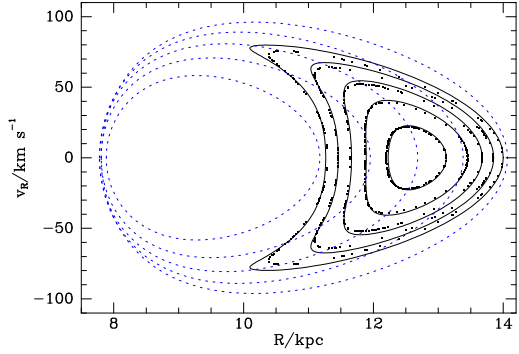}
\end{center}
\caption{A surface of section $\phi=z=0$ in a realistic barred Galaxy model.
The full black curves are   perturbatively constructed cross
sections through tori trapped by the bar's OLR. The points are consequents
on orbits started from a point on each of these curves. The dashed blue
curves show cross sections through the underlying axisymmetric tori.}\label{fig:sos2}
\end{figure}

Unless $\Phi_1$ is extremely small, sufficiently near the resonance $f_1$
becomes larger than $f_0$. This is problematic because the sign of $f_1$
fluctuates, so once $f_1$ exceeds $f_0$ the total DF $f_0+f_1$ is liable to
be negative, which is unphysical.

The linearised Boltzmann equation breaks down when orbits become trapped by a
resonance. Far from a resonance, the phase-space tori to which stars on
regular orbits are confined are merely distorted by $\Phi_1$. At a critical
distance from the resonance, the tori abruptly rearrange themselves into a
completely new pattern, and we say that the old tori/orbits have been trapped
by the resonance. Angle-action variables enable us to reduce this phenomenon
to a one-dimensional problem, closely analogous to the dynamics of a
pendulum. Untrapped orbits correspond to a pendulum that rotates fast enough
to pass top dead-centre and rotate always in the same sense, while trapped
orbits correspond to a pendulum that swings to and fro as in a clock. Trapped
orbits cannot be characterised by the same set of actions that characterise
untrapped orbits, but must be assigned an entirely new ``action of
libration'' in addition to two linear combinations of the original actions
\citep[e.g.][]{BinneyRes}. Consequently, for each family of resonantly
trapped orbits we require a new DF $f(\vJ)$ in addition to the familiar DF of
the untrapped orbits.

Fig.~\ref{fig:sos2} shows that when one uses the angle-action coordinates
that torus mapping provides and the enhanced pendulum equation that
\cite{Kaasalainen} introduced, one can obtain remarkably accurate analytic
models of trapped orbits. Specifically the figure shows a surface of section
$(R,v_R)$ at $\phi=0,z=0$ for motion in a realistic barred Galactic
potential. The full curves are cross sections through tori  trapped by the
OLR of a realistic barred Galaxy model.  These curves agree extremely well
with the lines of dots, which are consequents of orbits integrated from an
initial condition provided by one point on each curve. The blue dashed curves
are cross sections through the axisymmetric tori that underpin the
perturbative results.

{}
\end{document}